\documentclass[%
 reprint,
 amsmath,amssymb,
 aps,
]{revtex4-2}

\usepackage{graphicx}
\usepackage{dcolumn}
\usepackage{bm}

\usepackage{amsmath}
\usepackage{amssymb}
\usepackage{mathrsfs}
\usepackage{amsfonts}
\usepackage{amsthm}
\usepackage{color}
\usepackage{txfonts}
\usepackage[colorlinks=true,citecolor=blue,linkcolor=blue,urlcolor=blue,anchorcolor=blue]{hyperref}%
\usepackage{pifont}
\begin{document}

\preprint{APS/123-QED}

\title{Quantum entanglement and Einstein-Podolsky-Rosen steering \\ in magnon frequency comb}





\author{Qianjun Zheng$^1$}
\author{H. Y. Yuan$^2$}
\email[]{hyyuan@zju.edu.cn}
\author{Yunshan Cao$^1$}
\author{Peng Yan$^1$}
\email[]{yan@uestc.edu.cn}
\affiliation{$^1$School of Physics and State Key Laboratory of Electronic Thin Films and Integrated Devices, University of
Electronic Science and Technology of China, Chengdu 610054, China\\$^2$Institute for Advanced Study in Physics, Zhejiang University, Hangzhou 310027, China}
\begin{abstract}
Significant progress has been made for the emerging concept of magnon frequency comb (MFC) but mainly in the classical region. The quantum property of the comb structure is yet to be explored. Here we theoretically investigate the quantum fluctuations of frequency combs and demonstrate the continuous-variable quantum entanglement and Einstein-Podolsky-Rosen (EPR) steering between different teeth of MFC. Without loss of generality, we address this issue in a hybrid magnon-skyrmion system. We observe a strong two-mode squeezed entanglement and asymmetric steering between the sum- and difference-frequency magnon teeth mediated by the skyrmion that acts as an effective reservoir to cool the Bogoliubov mode delocalized over the first-order magnon pair in MFC. Our findings show the prominent quantum nature of MFC, which has the potential to be utilized in ultrafast quantum metrology and multi-task quantum information processing. 

\end{abstract}

\maketitle


\section{\label{sec:level1}Introduction}

Magnon frequency comb (MFC) is a spin-wave (SW, with its quantum called magnon) spectrum composed of a series of equidistant and coherent narrow spectral lines. The concept of MFC is attracting many recent attentions because of both fundamental interest and potential application in high-precision frequency measurement, ultra-sensitive detection, and ultra-fast magnon devices with the analogy of the optical frequency comb \cite{cundiff2003colloquium,picque2019frequency,fortier201920}. It has been shown that the MFC can be generated by means of various nonlinear effects, such as the three- and four-magnon scattering \cite{wang2021magnonic,hula2022spin}, magnetostriction \cite{xiong2023magnonic,liu2023generation}, magnon Kerr effect \cite{wang2023nonreciprocal,rao2023unveiling}, magnon-optical Brillouin scattering \cite{liu2022optomagnonic}, etc. The frequency range of coherent peaks in MFC has been extended from gigahertz (GHz) in ferromagnets to terahertz (THz) in antiferromagnets \cite{jin2023nonlinear,yao2023terahertz,zhang2024anisotropic}, and the controllable MFC bridging these two frequency domains is theoretically suggested in synthetic ferrimagnets as well \cite{liu2024design}. Xu \emph{et al.} \cite{xu2023magnonic} experimentally achieved a MFC with up to 20 comb teeth by driving a giant mechanical resonance with a strong pump field. Lately, Wang \emph{et al.} \cite{wang2024enhancement} reported a tunable and low-power MFC with more than 32 comb teeth by using exceptional points which enhance the nonlinear interaction between the pump-induced magnon mode and the Kittel mode. Liu \cite{liu2024dissipative} theoretically predicted an ultra-wideband MFC containing up to 400 comb teeth based on the dissipative coupling between the magnon and cavity photon, indicating that there exists still a large room for exploring MFC with dense teeth. It has been proposed that the nonreciprocal MFC can be generated in the dual-cavity magnonic system via the asymmetrical response of magnon-Kerr nonlinearity in two different microwave input directions \cite{wang2023nonreciprocal}. More recently, Liang \emph{et al.} \cite{liang2024asymmetric} demonstrated an asymmetric MFC through the nonlinear interaction between the skyrmion and chiral SW in a hybrid magnon-waveguide$|$skyrmion structure. 

Three-magnon scattering is the leading nonlinear effect in magnon dynamics, which includes both the confluence and splitting processes \cite{ordonez2009three}. However, it is normally weak in uniformly magnetized ferromagnets due to the weak dipolar interaction. Intriguingly, it is theoretically proposed that the MFC can be generated by enhancing the nonlinearity through the interaction between propagating magnons and topological magnetic textures, such as skyrmions \cite{wang2021magnonic}, bimerons \cite{zhang2024anisotropic}, vortices \cite{wang2022twisted}, and domain walls \cite{zhang2018eavesdropping,zhou2021spin}. It is worth noting that the magnetic skyrmion, a vortex-like structure with topological protection, holds great potential for future high-density information storage and robust spintronic devices owing to its advantages of nanoscale size, stable structure, convenient manipulation, and low energy consumption \cite{nagaosa2013topological,fert2017magnetic}. From another perspective, the nonlinear magnon-skyrmion scattering is inherently a quantum parametric oscillation process that describes the cascaded down-conversion and sum-frequency generation, which may imply interesting quantum effects of comb teeth.

Quantum entanglement is one of the most exotic features of quantum mechanics, which can be traced back to the Einstein-Podolsky-Rosen (EPR) paradox about the argument on the completeness of quantum theory \cite{einstein1935can}. In response to this paradox, Schr\"{o}dinger introduced entangled state and first proposed steering to describe the nonlocality of spooky action at a distance in EPR paradox \cite{schrodinger1935discussion}. In a system composed of two subsystems A and B, B can be controlled to the corresponding state by performing suitable measurements on A, which is called quantum steering \cite{uola2020quantum}. Different from the quantum entanglement, the EPR steering has more stringent correlation condition and exhibits an intrinsic asymmetry, or even unidirectionality, that is to say, one party of the two subsystems can control the quantum state of the other one, but not vice versa \cite{handchen2012observation,he2015classifying}. In the past few decades, quantum entanglement and steering as the key resources have played important roles in quantum computing \cite{madsen2022quantum}, quantum cryptography \cite{yin2020entanglement}, quantum sensing \cite{malia2022distributed}, quantum teleportation \cite{ren2017ground}, and quantum key distribution \cite{liao2017satellite,zhang2022device}. Although quantum correlations between magnons and various quasi-particles have been investigated \cite{li2018magnon,zhang2019quantum,zhang2019quantum,tan2019genuine,lachance2020entanglement,yuan2020steady,YUAN20221,hei2023enhanced,zhong2023nonreciprocal,zheng2023tutorial}, their manifestation among different magnon teeth in the emerging MFC is still an open issue.

In this work, we study the quantum correlations of spectral lines of MFC. Without loss of generality, we consider a coupled system consisting of a magnon waveguide and a magnetic skyrmion microresonator [see Fig. \ref{fig1}(a)]. The isolated magnetic skyrmion supporting a circular 180° domain wall can be treated as a microring resonator (the quality factor $Q\approx10^4$) that allows the magnonic whispering gallery mode (mWGM). When the mWGM driven by a microwave field enters the microresonator along the waveguide, it can be nonlinearly coupled to the skyrmion mode (e.g., skyrmion breathing, gyration, or skyrmion-wall circling modes), leading to the generation of MFC. We find that there are two separate parameter regimes, one of which has a finite threshold while the other has a divergent threshold, depending on the model parameters. When the driving power is below the threshold, only the mWGM is macroscopically occupied. In such a case, the mWGM can be parametrically approximated as a classical state and it is decoupled from the skyrmion microresonator. For the case above the threshold, all magnon modes would be excited but their solutions exhibit phase shifts, which is induced by the strong nonlinearity \cite{PhysRevB.107.L060401} and makes the determination of the phases challenging. In this study, we only consider the quantum fluctuations of MFC below the threshold by the standard linearization procedure. It is found that the entanglement between the skyrmion mode and the difference-frequency mode originating from their two-mode squeezing interaction can be transferred to the sum- and difference-frequency modes. Intriguingly, the strong entanglement and asymmetric steering between the two magnons are mediated by the skyrmion which acts as an engineered reservoir to cool the delocalized Bogoliubov mode through the linear beam-splitter interaction. We show that one party with larger effective magnon number steers the other one, and the steering direction can be flexibly controlled by the magnon dissipations. We interpret the underlying mechanism by constructing an effective Hamiltonian, and discuss parameters in tuning the quantum entanglement and EPR steering. In experiments, the comb lines close to the pumped magnon mode will be significantly excited, while magnon modes far from the pump remain unoccupied macroscopically. Therefore, the above quantum correlation analysis based on the two-mode squeezed model is well applicable to modes away from the pump.

This paper is organized as follows. In Sec. \ref{theoretical model}, we introduce the theoretical model to describe the nonlinear magnon-skyrmion interaction. The system Hamiltonian and the corresponding quantum Langevin equations (QLEs) are provided. In Sec. \ref{linear analysis}, we give the stationary semi-classical solutions and analyze the linearized quantum fluctuation dynamics. The bipartite entanglement and quantum steering criteria are shown in Sec. \ref{entanglement}. We then discuss the generation and manipulation of quantum correlation of the skyrmion and the magnon comb modes in Sec. \ref{result and discussion}. The verification of two-mode squeezed entanglement of the first-order magnon pair with respect to the reconstructed Wigner function and the experimental measurement scheme of quantum entanglement and steering are given in Sec. \ref{Verification and Measurement}. Conclusions are drawn in Sec. \ref{conclusion}.

\section{Theoretical model}\label{theoretical model}

\begin{figure}
\includegraphics[width=8cm]{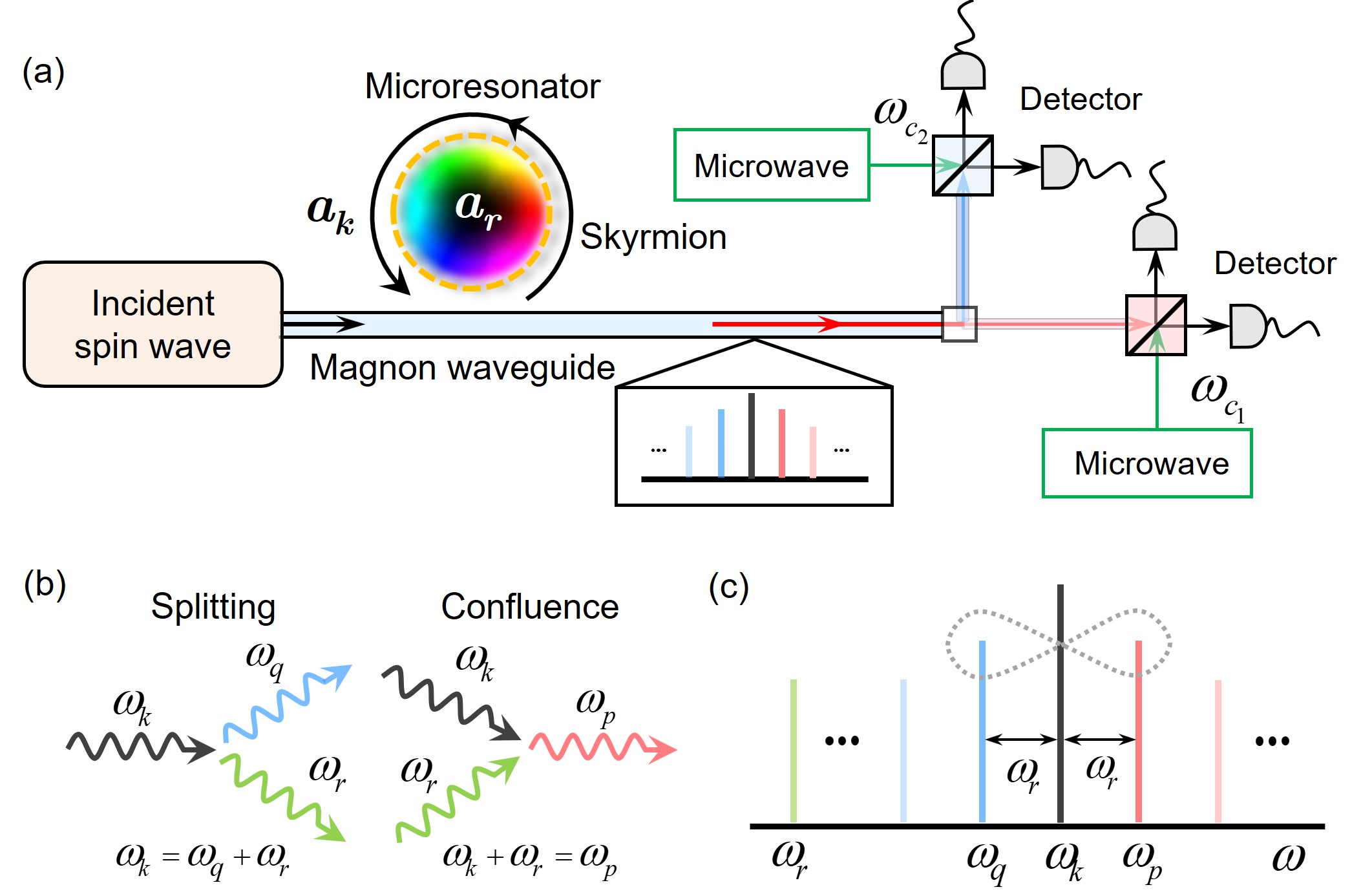}
\caption{\label{fig1} (a) Schematic illustration of the magnon-skyrmion hybrid system for the generation of MFC and the measurement scheme of the quantum entanglement and steering. The driven SW propagating in a magnon waveguide is incident into a skyrmion microresonator, simultaneously stimulating the mWGM $a_{k}$ with frequency $\omega_k$, which can be coupled with the skyrmion mode $a_{r}$ with frequency $\omega_r$ via the three-magnon process. Weak microwave fields with frequencies $\omega_{c_1,c_2}$ are sent to couple to the output MFC and the position and momentum of the magnon can be measured by homodyning the microwave field output. (b) Splitting and confluence processes of the nonlinear magnon-skyrmion scattering. (c) The spectrum of an ideal MFC is discrete, equally spaced, and wide-band, where the comb spacing is determined by the frequency of the skyrmion mode and different comb teeth can be entangled, labeled by the dotted figure-of-eight.}
\end{figure}

We consider a hybrid system composed of a magnon waveguide and a skyrmion microresonator that supports the mWGM, as shown in Fig. \ref{fig1}(a). The single magnetic skyrmion can be regarded as a microresonator because its magnetization rotates smoothly from the core down to the perimeter up, forming a circular 180$^{\circ}$ domain wall \cite{garcia2015narrow} that allows gapless SW excitations \cite{liang2024asymmetric}. When
the driving SW is injected into the skyrmion microresonator through the waveguide, a
mWGM with the same frequency is excited. It is then split into a skyrmion mode and a difference-frequency mode via the three-magnon process, and the confluence between the mWGM and the skyrmion mode generates the sum-frequency mode, as shown in Fig. \ref{fig1}(b). The chain-like nonlinear processes finally induce the MFC \cite{wang2021magnonic}. Here, we focus on the possible quantum correlation among the skyrmion ($\omega_{r}$) and lowest three MFC teeth ($\omega_{k}, \omega_{p}, \omega_{q}$), as depicted in Fig. \ref{fig1}(c). The Hamiltonian describing the nonlinear magnon-skyrmion interaction is written as ($\hbar=1$)%
\begin{align}
H=&\omega_{k}a_{k}^{\dagger}a_{k}+\omega_{r}a_{r}^{\dagger}a_{r}+\omega
_{p}a_{p}^{\dagger}a_{p}+\omega_{q}a_{q}^{\dagger}a_{q}\nonumber\\
&+g_{p}(a_{k}a_{r}a_{p}^{\dagger}+a_{k}^{\dagger}a_{r}^{\dagger}a_{p}%
)+g_{q}(a_{k}a_{r}^{\dagger}a_{q}^{\dagger}+a_{k}^{\dagger}a_{r}%
a_{q})\nonumber\\
&+iE(a_{k}^{\dagger}e^{-i\omega_{0}t}-a_{k}e^{i\omega_{0}%
t}),\label{eq1}%
\end{align}
where $\omega_{k}$ and $\omega_{r}$ are the resonant frequencies of the mWGM and the skyrmion mode, respectively, with $a_{k}\ (a_{k}^{\dagger})$ and $a_{r}\ (a_{r}^{\dagger})$ being their annihilation (creation) operators accordingly. $a_{p}\ (a_{p}^{\dagger})$ and $a_{q}\ (a_{q}^{\dagger})$ are the annihilation (creation) operators of the sum- and difference-frequency magnons, respectively, whose
frequencies satisfy $\omega_{p}=\omega_{k}+\omega_{r}$ and $\omega_{q}%
=\omega_{k}-\omega_{r}$. $g_{p}$ and $g_{q}$ are the nonlinear magnon-skyrmion coupling
strengths, which are considered as real numbers for simplicity. The incident SW is driven by the
microwave field with the frequency $\omega _{0}$ and the amplitude $E$ being the strength of the driving field.

In the rotating frame at the frequency $\omega_{0}$, the system Hamiltonian can be recast as%
\begin{align}
H  =&\Delta_{k}a_{k}^{\dagger}a_{k}+\omega_{r}a_{r}^{\dagger}a_{r}+\Delta
_{p}a_{p}^{\dagger}a_{p}+\Delta_{q}a_{q}^{\dagger}a_{q}\nonumber\\
  &  +g_{p}(a_{k}a_{r}a_{p}^{\dagger}+a_{k}^{\dagger}a_{r}^{\dagger}a_{p}%
)+g_{q}(a_{k}a_{r}^{\dagger}a_{q}^{\dagger}+a_{k}^{\dagger}a_{r}%
a_{q})\nonumber\\
  &  +iE(a_{k}^{\dagger}-a_{k}),\label{eq2}%
\end{align}
where $\Delta_{k(p,q)}=$\ $\omega_{k(p,q)}-$\ $\omega_{0}$.\ Considering the magnon dissipations and environmental noise, we derive the QLEs as follows%
\begin{align}
\dot{a}_{k}   =&-(i\Delta_{k}+\kappa_{k})a_{k}-ig_{p}a_{r}^{\dagger}%
a_{p}-ig_{q}a_{r}a_{q}+E\nonumber\\
&+\sqrt{2\kappa_{k}}a_{k}^{\text{in}},\nonumber\\
\dot{a}_{r}   =&-(i\omega_{r}+\kappa_{r})a_{r}-ig_{p}a_{k}^{\dagger}%
a_{p}-ig_{q}a_{k}a_{q}^{\dagger}+\sqrt{2\kappa_{r}}a_{r}^{\text{in}},\nonumber\\
\dot{a}_{p}   =&-(i\Delta_{p}+\kappa_{p})a_{p}-ig_{p}a_{k}a_{r}+\sqrt
{2\kappa_{p}}a_{p}^{\text{in}},\nonumber\\
\dot{a}_{q}   =&-(i\Delta_{q}+\kappa_{q})a_{q}-ig_{q}a_{k}a_{r}^{\dagger
}+\sqrt{2\kappa_{q}}a_{q}^{\text{in}},\label{eq3}%
\end{align}
where $\kappa_{j}$ $(j=k,r,p,q)$ and $a_{j}^{\text{in}}$ are the dissipation
coefficient and the input noise operator of the corresponding magnon mode, respectively.  Under the Markovian reservoir assumption, the input noise is characterized by the
zero-mean correlation functions: $\left\langle a_{j}^{\text{in}}(t)a_{j}^{\text{in}\dagger
}(t^{\prime})\right\rangle =[\bar{n}_{j}(\omega_{j})+1]\delta(t-t^{\prime})$,
and $\left\langle a_{j}^{\text{in}\dagger}(t)a_{j}^{\text{in}}(t^{\prime})\right\rangle
=\bar{n}_{j}(\omega_{j})\delta(t-t^{\prime})$. The equilibrium mean thermal
number of each mode is $\bar{n}_{j}(\omega_{j})=[\exp(\hbar\omega_{j}/k
_{B}T)-1]^{-1}$, where $T$ is the environmental temperature and $k_{B}$
is the Boltzmann constant.

\section{Linearized quantum fluctuation analysis}\label{linear analysis}
When the system is at the steady state, the dynamics can be linearized by expressing the operators as the sum of their expectation values and quantum fluctuations $a_{j}=\left\langle a_{j}%
\right\rangle +\delta a_{j}$. The mean value solutions could be obtained from a set of semi-classical
differential equations when $d\left\langle a_{j}\right\rangle /dt=0$. In these
cases we may use the linearized quantum fluctuation analysis to simplify the
physical model, and further study quantum correlation among different modes. 

Neglecting all fluctuations terms, the semi-classical equations for the magnon mean values are expressed as
\begin{align}
\langle\dot{a}_{k}\rangle &  =-(i\Delta_{k}+\kappa_{k})\langle a_{k}%
\rangle-ig_{p}\langle a_{r}^{\dagger}\rangle\langle a_{p}\rangle-ig_{q}\langle
a_{r}\rangle\langle a_{q}\rangle+E,\nonumber\\
\langle\dot{a}_{r}\rangle &  =-(i\omega_{r}+\kappa_{r})\langle a_{r}%
\rangle-ig_{p}\langle a_{k}^{\dagger}\rangle\langle a_{p}\rangle-ig_{q}\langle
a_{k}\rangle\langle a_{q}^{\dagger}\rangle,\nonumber\\
\langle\dot{a}_{p}\rangle &  =-(i\Delta_{p}+\kappa_{p})\langle a_{p}%
\rangle-ig_{p}\langle a_{k}\rangle\langle a_{r}\rangle,\nonumber\\
\langle\dot{a}_{q}\rangle &  =-(i\Delta_{q}+\kappa_{q})\langle a_{q}%
\rangle-ig_{q}\langle a_{k}\rangle\langle a_{r}^{\dagger}\rangle.\label{eq4}%
\end{align}
For convenience, we write the driving field and steady-state mean value of the system in complex numbers $E=\varepsilon e^{i\phi _{l}}$ and $\langle
a_{j}\rangle =A_{j}e^{i\phi _{j}}$. We find that the steady-state solutions are divided into two different classes depending on whether the threshold of the system driving field is finite or divergent.

If $g_{q}^{2}\kappa _{p}>g_{p}^{2}\kappa _{q}$,  we find a threshold field amplitude 
\begin{equation}
\varepsilon _{\text{th}}=\kappa _{k}\sqrt{\frac{\kappa _{r}\kappa _{p}\kappa
_{q}}{g_{q}^{2}\kappa _{p}-g_{p}^{2}\kappa _{q}}},  \label{eq5}
\end{equation}
with $\Delta _{k}=0$.  Below the threshold, only the mWGM matching the driving frequency $%
\omega _{0}$ is resonantly excited while other magnons will not be macroscopically occupied. In this circumstance, the steady-state solutions can be obtained 
\begin{equation}
\langle a_{k}\rangle =E/\kappa _{k},\langle a_{r}\rangle =\langle
a_{p}\rangle =\langle a_{q}\rangle =0.  \label{eq6}
\end{equation}
When the pumping amplitude $\varepsilon$ is above the threshold,  there exist another steady-state solution of the mWGM, skyrmion, sum- and difference-frequency magnon modes as
\begin{align}
\langle a_{k}\rangle & =\sqrt{\frac{\kappa _{r}\kappa _{p}\kappa _{q}}{%
g_{q}^{2}\kappa _{p}-g_{p}^{2}\kappa _{q}}}e^{i\phi _{k}},\text{ }  \notag \\
\langle a_{r}\rangle & =\sqrt{\frac{(\varepsilon -\varepsilon _{\text{th}%
})\kappa _{k}\kappa _{p}\kappa _{q}}{\varepsilon _{\text{th}%
}(g_{q}^{2}\kappa _{p}+g_{p}^{2}\kappa _{q})}}e^{i\phi _{r}},  \notag \\
\langle a_{p}\rangle & =g_{p}\kappa _{q}\sqrt{\frac{(\varepsilon
-\varepsilon _{\text{th}})\kappa _{k}\kappa _{r}}{\varepsilon _{\text{th}%
}(g_{q}^{4}\kappa _{p}^{2}-g_{p}^{4}\kappa _{q}^{2})}}e^{i\phi _{p}},  \notag
\\
\langle a_{q}\rangle & =g_{q}\kappa _{p}\sqrt{\frac{(\varepsilon
-\varepsilon _{\text{th}})\kappa _{k}\kappa _{r}}{\varepsilon _{\text{th}%
}(g_{q}^{4}\kappa _{p}^{2}-g_{p}^{4}\kappa _{q}^{2})}}e^{i\phi _{q}},
\label{7}
\end{align}
where the phases satisfy $\phi _{k}=\phi _{l},\phi
_{p}+\phi _{q}=2\phi _{k}+\pi ,\phi _{p}-\phi _{q}=2\phi _{r}$.
The only phase we know here is the phase of the driving field $E$. If we take $E$ as a real number, $\langle a_{k}\rangle $ will also be a real number. But other phases $\phi _{r,p,q}$ are not fixed, which indicates phase shifts of these solutions above the threshold \cite{PhysRevB.107.L060401}, that is, they are uncertain. 

If $g_{q}^{2}\kappa _{p}\leq g_{p}^{2}\kappa _{q}$, the threshold diverges, which implys that no matter how strong the pump field is, other magnon modes will not be macroscopically occupied. The expressions of the stationary solutions are then the same as Eq. (\ref{eq6}).

In what follows, we shall only investigate the quantum correlation for model parameters allowing the steady-state solutions (\ref{eq6}). Based on Eq. (\ref{eq3}), we can linearize the fluctuations around the steady state, and obtain
\begin{align}
\delta\dot{a}_{k}   =&-(i\Delta_{k}+\kappa_{k})\delta a_{k}-ig_{p}\langle
a_{r}^{\dagger}\rangle\delta a_{p}-ig_{p}\langle a_{p}\rangle\delta
a_{r}^{\dagger}\nonumber\\
&-ig_{q}\langle a_{r}\rangle\delta a_{q}-ig_{q}\langle a_{q}\rangle\delta a_{r}+\sqrt{2\kappa_{k}}\delta a_{k}^{\text{in}},\nonumber\\
\delta\dot{a}_{r}   =&-(i\omega_{r}+\kappa_{r})\delta a_{r}-ig_{p}\langle
a_{k}^{\dagger}\rangle\delta a_{p}-ig_{p}\langle a_{p}\rangle\delta
a_{k}^{\dagger}\nonumber\\
&-ig_{q}\langle a_{k}\rangle\delta a_{q}^{\dagger}-ig_{q}\langle a_{q}^{\dagger}\rangle\delta a_{k}+\sqrt{2\kappa_{r}}\delta
a_{r}^{\text{in}},\nonumber\\
\delta\dot{a}_{p}   =&-(i\Delta_{p}+\kappa_{p})\delta a_{p}-ig_{p}\langle
a_{k}\rangle\delta a_{r}-ig_{p}\langle a_{r}\rangle\delta a_{k}\nonumber\\
&+\sqrt{2\kappa_{p}}\delta a_{p}^{\text{in}},\nonumber\\
\delta\dot{a}_{q}   =&-(i\Delta_{q}+\kappa_{q})\delta a_{q}-ig_{q}\langle
a_{k}\rangle\delta a_{r}^{\dagger}-ig_{q}\langle
a_{r}^{\dagger}\rangle\delta
a_{k}\nonumber\\
&+\sqrt{2\kappa_{q}}\delta a_{q}^{\text{in}}.\label{eq8}%
\end{align}
By introducing the magnons position and momentum quadrature fluctuation
operators $X_{j}=(\delta a_{j}+\delta a_{j}^{\dagger})/\sqrt{2}$,
$Y_{j}=i(\delta a_{j}^{\dagger}-\delta a_{j})/\sqrt{2}$, and the input noise
quadrature operators $X_{j}^{\text{in}}=(\delta a_{j}^{\text{in}}+\delta a_{j}^{\text{in}\dagger
})/\sqrt{2}$, $Y_{j}^{\text{in}}=i(\delta a_{j}^{\text{in}\dagger}-\delta a_{j}^{\text{in}}%
)/\sqrt{2}$, the linearized QLEs describing the quantum fluctuations can be recast as
\begin{equation}
\mathbf{\dot{u}}(t)=M\mathbf{u}(t)+\mathbf{n}(t),\newline\label{eq9}%
\end{equation}
where the vector of quadrature operators $\mathbf{u}(t)=[\delta X_{k}%
(t),\delta Y_{k}(t),\delta X_{r}(t),\delta Y_{r}(t),\delta X_{p}(t),\delta
Y_{p}(t),\delta X_{q}(t),
\delta Y_{q}(t)]^{\text{T}}$, the vector of noise operators
$\mathbf{n}(t)=[\sqrt{2\kappa_{k}}X_{k}^{\text{in}},\sqrt{2\kappa_{k}}Y_{k}%
^{\text{in}},\sqrt{2\kappa_{r}}X_{r}^{\text{in}},\sqrt{2\kappa_{r}}Y_{r}^{\text{in}},\sqrt
{2\kappa_{p}}X_{p}^{\text{in}},
\sqrt{2\kappa_{p}}Y_{p}^{\text{in}},\\
\sqrt{2\kappa_{q}}%
X_{q}^{\text{in}},\sqrt{2\kappa_{q}}Y_{q}^{\text{in}}]^{\text{T}}$, and $M$ is the drift matrix. If
the real parts of all eigenvalues of $M$ are negative, the system is stable
and reaches a steady state, known as the Routh-Hurwitz
criterion \cite{dejesus1987routh}. Because the dynamics of quadrature components fluctuations are linearized and the input
noises are Gaussian, the system finally evolves into a zero-mean continuous
variable (CV) Gaussian state that can be characterized by a $8\times8$
covariance matrix (CM) $V$, where the matrix elements are defined as
$V_{kl}=\left\langle \mathbf{u}_{k}(\infty)\mathbf{u}_{l}(\infty
)+\mathbf{u}_{l}(\infty)\mathbf{u}_{k}(\infty)\right\rangle /2$
$(k,l=1,2,\ldots8)$. From Eq. (\ref{eq9}), the formal solution of the linearized QLEs
is given by $\mathbf{u}(t)=\mathbf{\Lambda}(t)\mathbf{u}(0)+\int_{0}%
^{t}ds\mathbf{\Lambda}(s)\mathbf{n}(t-s)$ with $\mathbf{\Lambda}(t)=\exp(Mt)$.
In the steady state $t\rightarrow\infty$, one acquires $\mathbf{\Lambda
}(\infty)=0$ and $\mathbf{u}_{k}(\infty)=\int_{0}^{\infty}ds\sum
\nolimits_{i}\mathbf{\Lambda}_{ki}(s)\mathbf{n}(t-s)$. Based on the fact that
the eight components of $\mathbf{n}(t)$ are uncorrelated with each other, one
gets $V=\int_{0}^{\infty}ds\mathbf{\Lambda}(s)D\mathbf{\Lambda}^{\text{T}}(s)$,  where $D$ is the diffusion matrix defined as $D_{kl}\delta(s-s^{\prime})=$
$\left\langle \mathbf{n}_{k}(s)\mathbf{n}_{l}(s^{\prime})+\mathbf{n}%
_{l}(s^{\prime})\mathbf{n}_{k}(s)\right\rangle /2$. When the system stability
conditions are fulfilled, the steady-state CM $V$ can be obtained by solving
the Lyapunov equation \cite{vitali2007optomechanical}
\begin{equation}
MV+VM^{\text{T}}=-D.\label{eq10}%
\end{equation}
The CM $V$ is a real and symmetric matrix and must satisfy the
Robertson-Schr\"{o}dinger uncertainty relation \cite{simon1994quantum,weedbrook2012gaussian} $V+i\Omega_{n}>0$, which is the necessary and sufficient condition for the Gaussian state to be physical and implys the positive definiteness $V>0$. $\Omega_{n}=\oplus_{j=1}^{n}%
i\sigma_{y}$\ is the standard symplectic form ($\sigma_{y}$ is the $y$-Pauli matrix). With the CM $V$, one can analyze the entanglement properties among magnons with different frequencies in MFC at the steady state.

\section{Entanglement and steering criteria}\label{entanglement}
To quantitatively study the quantum correlation of magnons in
the multipartite CV system, we adopt the logarithmic negativity $E_{N}$ \cite{adesso2004extremal} to
calculate the bipartite entanglement, which is defined as%
\begin{equation}
E_{N}=\max\Big[0,-\ln(2v)\Big],\label{eq11}%
\end{equation}
where $v=$\ $\sqrt{\sum(\tilde{V})-\left[  \sum(\tilde{V})^{2}-4\det\tilde
{V}\right]  ^{1/2}}/\sqrt{2}$ is the minimal symplectic eigenvalue of the
partial transpose of a reduced $4\times4$ CM $\tilde{V}$ with $\sum(\tilde
{V})\equiv\det (V_{1})+\det (V_{2})-2\det (V_{12})$. The reduced CM $\tilde{V}$ of
two selected modes under consideration can be obtained by tracing out the rows
and columns of the uninteresting modes in $V$, which is given by%
\begin{equation}
\tilde{V}=\left(
\begin{array}
[c]{cc}%
V_{1} & V_{12}\\
V_{12}^{\text{T}} & V_{2}%
\end{array}
\right)  ,\label{eq12}%
\end{equation}
where $V_{1}$ and $V_{2}$ are $2\times2$\ block matrices corresponding to
modes $1$ and $2$, respectively. If $E_{N}>0$, namely, $v<1/2$, the considered
bipartition are entangled, which is equivalent to the Peres-Horodecki
criterion for certifying bipartite entanglement of Gaussian states \cite{simon2000peres}, and the
larger $E_{N}$\ the higher the degree of entanglement.\ Besides, the Gaussian
quantum steering can be calculated quantitatively by \cite{kogias2015quantification}
\begin{align}
S_{12} &  =\max\left\{  0,S(2V_{1})-S(2\tilde{V})\right\}  ,\nonumber\\
S_{21} &  =\max\left\{  0,S(2V_{2})-S(2\tilde{V})\right\}  ,\label{eq13}%
\end{align}
where $S(\sigma)=\left[  \ln\det(\sigma)\right]  /2$. A nonzero
$S_{12}>0\ (S_{21}>0)$ denotes that the bipartite Gaussian state is steerable
from mode $1\ (2)$ to mode $2\ (1)$ by applying Gaussian measurements on mode
$1\ (2)$, and the larger of its value denotes the stronger Gaussian
steerability. Moreover, the directionality of quantum steering could be better
analyzed by introducing the effective magnon number, which can be calculated
through the diagonal elements of the CM $V$ from
\begin{equation}
N_{j}=\left[  \left\langle \delta X_{j}^{2}\right\rangle +\left\langle \delta
Y_{j}^{2}\right\rangle -1\right]  /2,\label{eq14}%
\end{equation}
with $j$ labeling the magnon mode.

\section{Results and Discussion}\label{result and discussion}
In this section, we will demonstrate the quantum entanglement and EPR steering among the skyrmion mode and teeth of MFC. When the hybrid magnon-skyrmion system is resonantly driven with $\Delta_{k}=0$, the solution of $\langle a_{k}\rangle$ is given by a real number $E/\kappa_{k}$ by assuming $\phi_{k}=0$. From Eq. (\ref{eq8}), the QLEs of fluctuations are reduced to
\begin{align}
\delta\dot{a}_{k}   =&-\kappa_{k}\delta a_{k}+\sqrt{2\kappa_{k}%
}\delta a_{k}^{\text{in}},\nonumber\\
\delta\dot{a}_{r}   =&-(i\omega_{r}+\kappa_{r})\delta a_{r}-iG_{p}\delta a_{p}-iG_{q}\delta a_{q}^{\dagger}\nonumber+\sqrt{2\kappa_{r}}\delta a_{r}^{\text{in}},\nonumber\\
\delta\dot{a}_{p}   =&-(i\omega_{r}+\kappa_{p})\delta a_{p}-iG_{p}\delta a_{r}+\sqrt{2\kappa_{p}}\delta a_{p}^{\text{in}},\nonumber\\
\delta\dot{a}_{q}   =&(i\omega_{r}-\kappa_{q})\delta a_{q}-iG_{q}\delta a_{r}^{\dagger}+\sqrt{2\kappa_{q}}\delta a_{q}%
^{\text{in}},\label{eq15}%
\end{align}
where $G_{p}=g_{p}\langle a_{k}\rangle$ and $G_{q}=g_{q}\langle a_{k}\rangle$
are effective coupling strengths describing the confluence and splitting processes, respectively. Equations (\ref{eq15}) show that the mWGM is decoupled from the rest system, so we will focus on the fluctuation dynamics of the skyrmion, the sum- and difference-frequency magnons next. The corresponding drift matrix $M$ is given by
\begin{equation}
M=\left(
\begin{array}
[c]{cccccc}%
-\kappa_{r} & \omega_{r} & 0 & G_{p} & 0 & -G_{q}\\
-\omega_{r} & -\kappa_{r} & -G_{p} & 0 & -G_{q} & 0\\
0 & G_{p} & -\kappa_{p} & \omega_{r} & 0 & 0\\
-G_{p} & 0 & -\omega_{r} & -\kappa_{p} & 0 & 0\\
0 & -G_{q} & 0 & 0 & -\kappa_{q} & -\omega_{r}\\
-G_{q} & 0 & 0 & 0 & \omega_{r} & -\kappa_{q}%
\end{array}
\right)  .\label{eq16}%
\end{equation}
Correspondingly, $\mathbf{u}(t)=[\delta X_{r}(t),\delta Y_{r}(t),\delta X_{p}(t),
\delta Y_{p}(t),
\delta X_{q}(t),\\
\delta Y_{q}(t)]^{\text{T}}$, $\mathbf{n}(t)=[\sqrt{2\kappa_{r}}%
X_{r}^{\text{in}},\sqrt{2\kappa_{r}}Y_{r}^{\text{in}},\sqrt{2\kappa_{p}}X_{p}^{\text{in}},
\sqrt{2\kappa_{p}}Y_{p}^{\text{in}},\\
\sqrt{2\kappa_{q}}X_{q}^{\text{in}},\sqrt{2\kappa_{q}%
}Y_{q}^{\text{in}}]^{\text{T}},$ and $D=\text{diag}[\kappa_{r}(2\bar{n}_{a}+1),\kappa_{r}(2\bar{n}%
_{a}+1),\kappa_{p}(2\bar{n}_{p}+1),\kappa_{p}(2\bar{n}_{p}+1),\kappa_{q}%
(2\bar{n}_{q}+1),\kappa_{q}(2\bar{n}_{q}+1)]$. We can obtain the effective Hamiltonian for the quantum fluctuation operators as follows
\begin{align}
H_{\text{eff}}   =&\omega_{r}(\delta a_{r}^{\dagger}\delta a_{r}+\delta
a_{p}^{\dagger}\delta a_{p}-\delta a_{q}^{\dagger}\delta a_{q})+G_{q}(\delta a_{r}^{\dagger}\delta a_{q}^{\dagger}+\delta a_{r}\delta
a_{q}) \nonumber\\
&+G_{p}(\delta a_{r}^{\dagger}\delta a_{p}+\delta a_{r}\delta a_{p}^{\dagger
}).\label{eq17}%
\end{align}

\begin{figure}
\includegraphics[width=9cm]{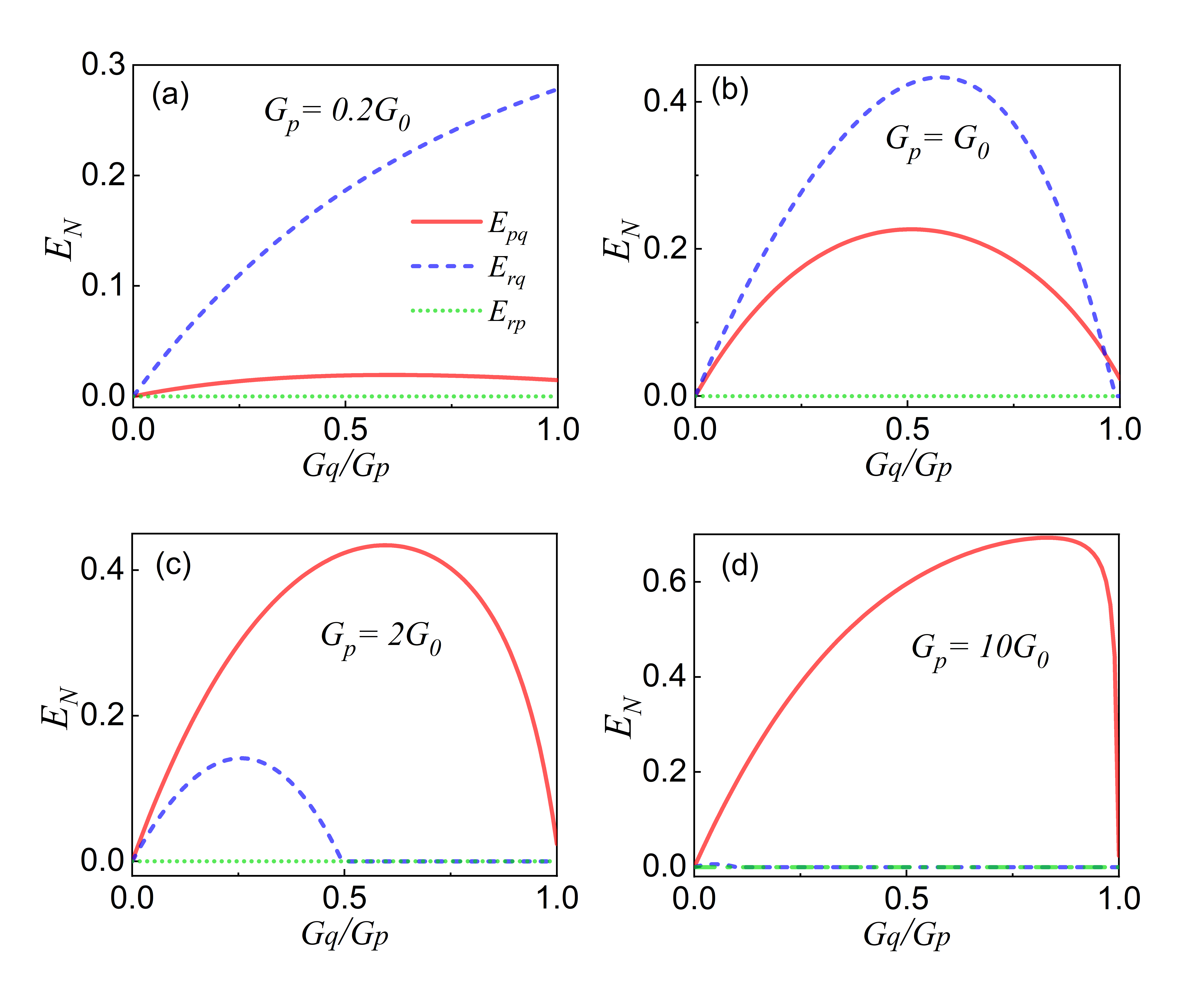}
\caption{\label{fig2} Logarithmic negativity of bipartite entanglements $E_{rp},E_{rq}$ and $E_{pq}$ versus the ratio of coupling strengths $G_{q}/G_{p}$ for (a) $G_{p}=0.2G_{0}$ , (b) $G_{p}=G_{0}$, (c) $G_{p}=2G_{0}$, and (d) $G_{p}=10G_{0}$ with $G_{0}/2\pi=15$ MHz.}
\end{figure}

To calculate the steady-state bipartite entanglements among different magnon modes, we adopt the following parameters \cite{wang2021magnonic}: $\omega_{k}/2\pi=$ 80 GHz, $\omega_{r}/2\pi=$ 8 GHz, $\omega_{p}/2\pi=$ 88 GHz, $\omega
_{q}/2\pi=$ 72 GHz, $\kappa_{p}/2\pi=\kappa_{q}/2\pi=$
10 MHz, $k_{r}/2\pi=$ 1 MHz, and $T=$ 20 mK. Figure \ref{fig2} show the logarithmic
negativity of the three bipartite cases versus the effective coupling ratio
$G_{q}/G_{p}$ for $G_{p}=0.2G_{0}$, $G_{p}=G_{0}$, $G_{p}=2G_{0}$, and $G_{p}=20G_{0}$ with $G_{0}/2\pi=$ 15 MHz, respectively. From Fig. \ref{fig2}(a), it can be found that the skyrmion mode and the difference-frequency mode are
entangled ($E_{rq}>0$), and the entanglement degree increases with the enhancement of the coupling of the magnon splitting process $G_{q}$. The weak entanglement between $a_{p}$ and $a_{q}$ also exists, but the skyrmion and the sum-frequency magnon are disentangled
($E_{rp}=0$). The entanglement between $a_{p}$ and $a_{q}$ is significantly enhanced for a stronger $G_{p}$, as illustrated in Fig. \ref{fig2}(b). And it can be seen that the entanglement between the skyrmion and the difference-frequency magnon is nonmonotonic with the increase of the coupling ratio $G_{q}$, and it disappears with the parameter $G_{q}=G_{p}$ (dashed blue curve). When the effective coupling $G_{p}$ is enhanced to $G_{p}=2G_{0}$, the degree of the entanglement between two magnons $E_{pq}$ is larger than that of the skyrmion-magnon entanglement $E_{rq}$ in Fig. \ref{fig2}(c). And it is obvious that the entanglement $E_{rq}$ only exists in the parameter interval $G_{q}<0.5G_{p}$. From Fig. \ref{fig2}(d), we can see that there is a strong entanglement between $a_{p}$ and $a_{q}$, while the skyrmion mode ($a_{r}$) and the magnon mode ($a_{q}$) can hardly be entangled for a very large effective coupling $G_{p}=10G_{0}$. Because the effective coupling strengths of the magnon-skyrmion interaction are directly proportional to the driving power of microwave field, it is not difficult to enhance $G_{p(q)}$ by 1$\sim$2 orders of magnitude. It is evident that this system has a sort of entanglement transfer (or sharing). One can interpret the physical mechanism via the effective Hamiltonian. With respect to the rotating frame at $\omega_{r}$, the effective Hamiltonian can be written as $H_{\text{eff}}^{\prime}=G_{q}(\delta
a_{r}^{\dagger}\delta a_{q}^{\dagger}+\delta a_{r}\delta a_{q})+G_{p}(\delta
a_{r}^{\dagger}\delta a_{p}+\delta a_{r}\delta a_{p}^{\dagger})$. We know that the first term is the parametric down-conversion (two-mode squeezed) coupling, which can entangle the skyrmion and the difference-frequency magnon. The second term is the linear beam-splitter coupling that generally does not create entanglement, but the entanglement can be partially transferred by means of such state exchange interaction, which leads to the entanglement between the sum- and difference-frequency magnon modes.

\begin{figure}
\includegraphics[width=9cm]{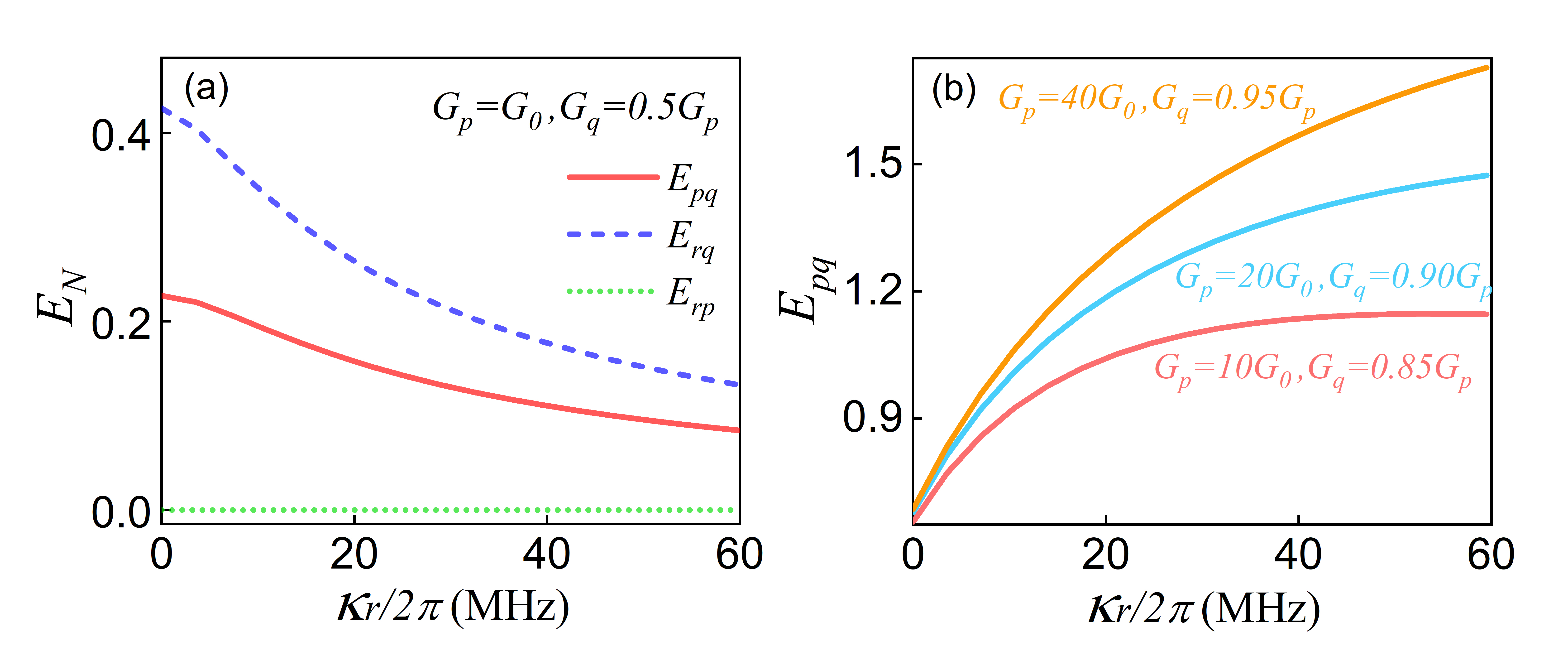}
\caption{\label{fig3}(a) Logarithmic negativity of bipartite entanglements
$E_{rp},E_{rq}$ and $E_{pq}$ versus the dissipation rate of skyrmion
$\kappa_{r}$ with $G_{p}=G_{0}$, and $G_{q}=0.5G_{p}$. (b) Bipartite
entanglement $E_{pq}$ versus the dissipation rate of skyrmion $\kappa_{r}$ for different effective coupling strengths. Other parameters are the same as those in Fig. \ref{fig2}.}
\end{figure}

Figure \ref{fig3} reveals the influence of the skyrmion dissipation on bipartite entanglements for different effective coupling strength $G_{p}$ with the optimal coupling ratio $G_{q}/G_{p}$ that corresponds to the maximum entanglement value. From Fig. \ref{fig3}(a), we observe that the steady-state entanglement gradually decreases with the increase of dissipation, showing a destructive role played by the skyrmion dissipation in the entanglement. However, a large enough dissipation would significantly enhance the entanglement of two magnon modes $a_{p}$ and $a_{q}$ with a strong coupling $G_{p}=10G_{0}$, as shown in Fig. \ref{fig3}(b). And the stronger entanglement can be obtained under the greater effective coupling with the optimized coupling ratio. Here, the generation of entanglement between two magnons under the condition of strong magnon-skyrmion interaction satisfies the mechanism of reservoir engineering \cite{wang2013reservoir}. To demonstrate this point, we introduce two delocalized Bogoliubov modes $\beta_{1}$\ and $\beta_{2}$ 
\begin{align}
\beta_{1} &  =a_{q}\cosh\xi+a_{p}^{\dagger}\sinh \xi=S(\xi)a_{q}S^{\dagger
}(\xi),\nonumber\\
\beta_{2} &  =a_{p}\cosh\xi+a_{q}^{\dagger}\sinh \xi=S(\xi)a_{p}S^{\dagger
}(\xi),\label{eq19}%
\end{align}
where $S(\xi)=\exp\big[\xi(a_{q}a_{p}-a_{q}^{\dagger}a_{p}^{\dagger})\big]$ is the
two-mode squeezing operator with the effective squeezing parameter
$\xi=\operatorname{arctanh}(G_{q}/G_{p})$. The system Hamiltonian (\ref{eq17}) can be rewritten
as ($\delta$ is omitted) $H_{B}=\omega_{r}(a_{r}^{\dagger}a_{r}+
\beta_{2}^{\dagger}\beta_{2}-\beta_{1}^{\dagger}\beta_{1})+\tilde{G}(a_{r}^{\dagger}\beta_{2}+a_{r}\beta_{2}^{\dagger})$ with
$\tilde{G}=\sqrt{G_{p}^{2}-G_{q}^{2}}$. It turns out that the mode
$\beta_{1}$ completely decouples with the skyrmion, and the mode $\beta_{2}$
has a beam-splitter type interaction with the skyrmion mode $a_{r}$. The squeezing operator $S(\xi)$ acts on the two-mode vacuum state $|0_{p},0_{q}\rangle$ and generates the two-mode squeezed vacuum state
$|\xi\rangle=S(\xi)|0_{p},0_{q}\rangle$, which is the joint vacuum state of
$\beta_{1}$\ and $\beta_{2}$ and is also a two-mode entanglement state. Therefore, the entanglement can be obtained by cooling the Bogoliubov modes
$\beta_{1}$\ and $\beta_{2}$ to their ground state. In our work, only
the $\beta_{2}$ mode is cooled by skyrmion that can be regarded as an
engineered reservoir via the beam-splitter interaction, but the strong
entanglement between $a_{p}$ and $a_{q}$ could still be realized. The cooling process is considered to be the predominant absorption of the Bogoliubov $\beta_{2}$ mode, which is eventually dissipated into the environment through the skyrmion leakage. The underlying state transition
between the states of the skyrmion and Bogoliubov $\beta_{2}$ modes is
$|n_{s},m_{\beta_{2}}\rangle\leftrightarrow|n_{s}+1,m_{\beta_{2}}-1\rangle$
controlled by the interaction term in $H_B$, yet the dissipation process of the skyrmion does not
change the Bogoliubov $\beta_{2}$ mode state with the transition
$|n_{s}+1,m_{\beta_{2}}-1\rangle\rightarrow|n_{s},m_{\beta_{2}}-1\rangle$,
where the skyrmion carries away the gained energy from the Bogoliubov
$\beta_{2}$ mode, leading to its cooling. Similar physical mechanism has been
utilized in hybrid cavity optomechanical and cavity magnonic systems to generate and manipulate the strong quantum correlation \cite{liao2018reservoir,guo2023manipulation,xie2023stationary,liu2023reservoir,shang2024generation}.

\begin{figure}
\includegraphics[width=9cm]{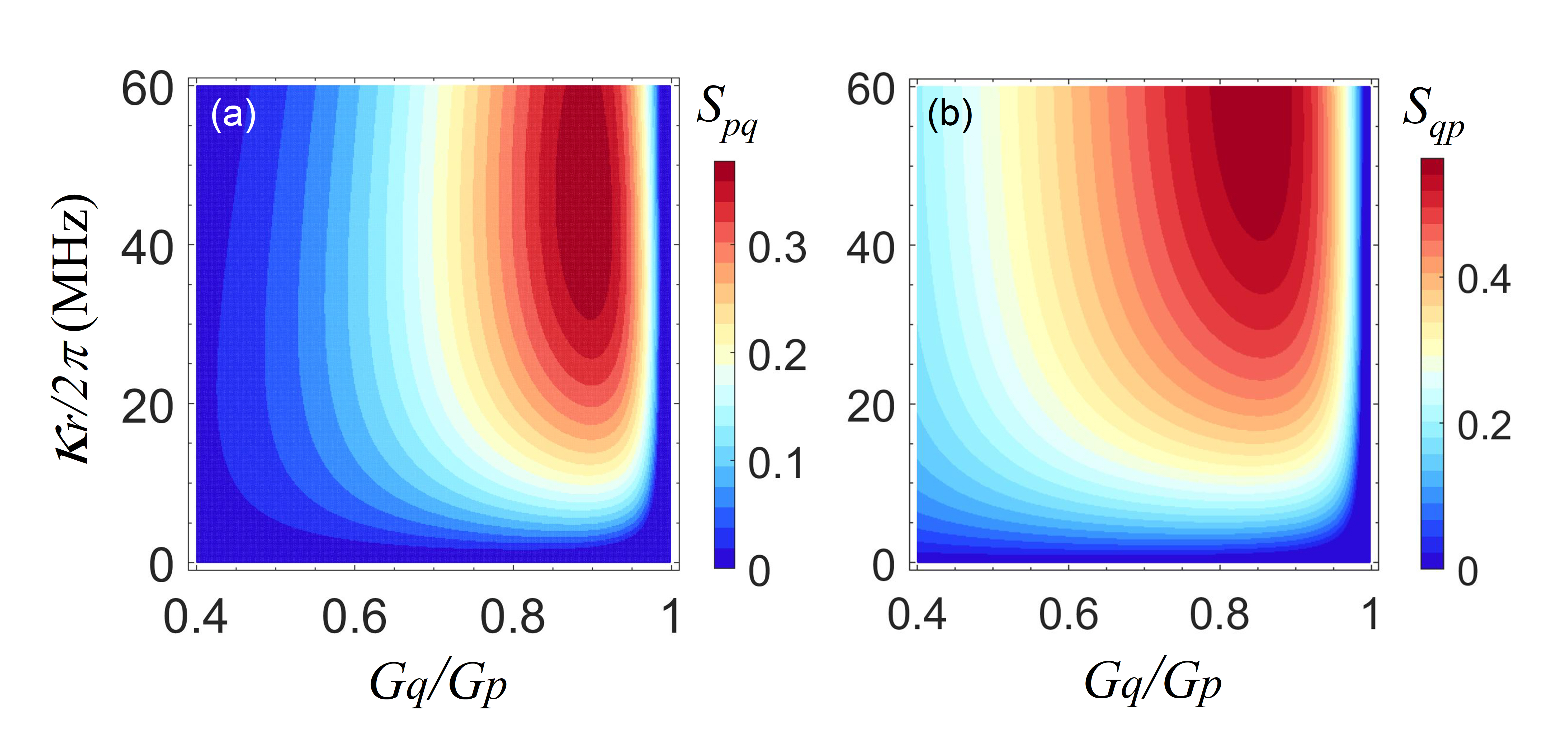}
\caption{\label{fig4} Density plot of the asymmetric EPR steering (a) $S_{pq}$ and (b) $S_{qp}$ versus the ratio of coupling strengths $G_{q}/G_{p}$ and the dissipation rate of skyrmion focus on with $G_{p}=10G_{0}$ and $\kappa_{r}/2\pi=40$ MHz.}
\end{figure}

As a strict subset of quantum entanglement, the EPR steering plays an important role in one side device-independent quantum key distribution because of its inherent directionality. Below we discuss the EPR steering between magnon modes $a_{p}$ and $a_{q}$. In Fig. \ref{fig4}, we can see that the steering can be strengthened by increasing the skyrmion dissipation $\kappa_{r}$, whereas they first increase and then decrease with the increase of $G_{q}$, and this nonmonotonic trend is similar to that of the entanglement curve in Fig. \ref{fig2}(d). On the one hand, enhancing $G_{q}$ for a fixed $G_{p}$ increases the squeezing parameter $\xi$, which leads to the enhancement of entanglement. On the other hand, the occupancies of the Bogoliubov modes $\left\langle \beta_{1(2)}^{\dagger}\beta_{1(2)}\right\rangle
=\cosh^{2}\xi\left\langle a_{q(p)}^{\dagger}a_{q(p)}\right\rangle +\sinh
^{2}\xi\left\langle a_{p(q)}^{\dagger}a_{p(q)}\right\rangle +\sinh \xi\cosh
\xi\left\langle a_{q}^{\dagger}a_{p}^{\dagger}+a_{q}a_{p}\right\rangle $ become exponentially large with the increase of $\xi$ where $\left\langle
a_{p(q)}^{\dagger}a_{p(q)}\right\rangle $ is the thermal magnon number, and the cooling effect of the skyrmion is suppressed due to the effective coupling $\tilde{G}$ decreasing with increasing $G_{q}$, thus the quantum correlation is weakened. Comparing Fig. \ref{fig4}(a) with Fig. \ref{fig4}(b), it can be seen that the asymmetric steering between the two magnons emerges. We observe that the bipartite steering parameter $S_{qp}$ is always larger than $S_{pq}$. It indicates that applying Gaussian measurements on $a_{p}$ or $a_{q}$ mode could bidirectionally steer the state of the other mode, yet the steerability of $a_{q}$ is stronger than that of $a_{p}$. It is worth noting that the asymmetric steering is intrinsic in our system, which does not require imposed unbalanced losses or noises on two parties at the expense of reduced steerability.

\begin{figure}
\includegraphics[width=9cm]{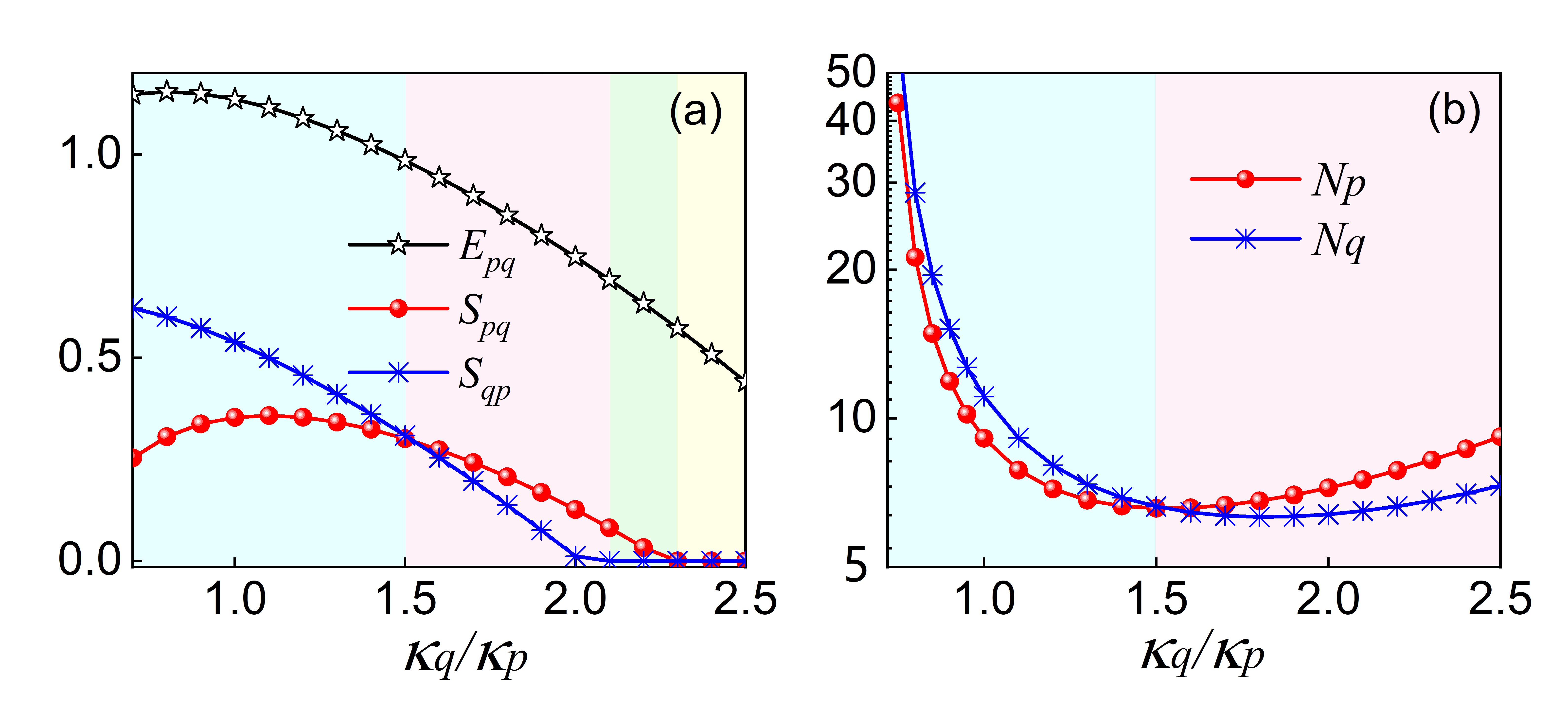}
\caption{\label{fig5} (a) Bipartite entanglement $E_{pq}$, bipartite steering $S_{pq}$, and
$S_{qp}$ versus the ratio of dissipation rate $\kappa_{q}/\kappa_{p}$ with
$G_{p}=10G_{0}$ MHz, $G_{q}=0.85G_{p}$, $\kappa_{p}/2\pi=10$ MHz, and $\kappa_{r}/2\pi=40$ MHz. (b) The effective
magnon numbers $N_{p}$ and $N_{q}$ versus the ratio of dissipation rate
$\kappa_{q}/\kappa_{p}$.}
\end{figure}

Additionally, the steering direction could be performed in a controllable manner by changing the dissipation rates of $a_{p}$ and $a_{q}$. The asymmetric external environments of two magnons can cause asymmetric quantum correlation that is beneficial for achieving the one-way steering. The stationary entanglement and steering of two magnons as functions of the dissipation ratio $\kappa_{q}/\kappa_{p}$ with a fixed $\kappa_{p}/2\pi=10$ MHz are displayed on Fig. \ref{fig5}(a), and the regions of two-way (blue and red), one-way (green), and no-way (yellow) EPR steering are respectively indicated by different colors. It is found that the correlation of two-magnon entanglement is always larger than the quantum steering for the same parameters regime ($E_{pq}>S_{pq},S_{qp}$). Both the entanglement and steering decrease with the increase of $\kappa_{q}/\kappa_{p}$, which indicates that the magnon damping is destructive to the quantum correlation between two magnons. In the parameter regime $\kappa_{q}/\kappa_{p}<2$, there is a bidirectional steering, in which the steering directions of the two magnons are competitive. With the increase of $\kappa_{q}/\kappa_{p}$, $S_{qp}$ is larger than $S_{pq}$, indicating a stronger steerable ability from the $q$-mode to $p$-mode with the parameters regime $\kappa_{q}/\kappa_{p}<1.5$ (blue region). However, when the dissipation ratio falls into the region $1.5<\kappa_{q}/\kappa_{p}<2.1$, the steering direction is reversed and the steerability from the $p$-mode to the $q$-mode is dominant ($S_{qp}<S_{pq}$, red region). An intuitive picture is as follows: the large $\kappa_{q}/\kappa_{p}$ means that the interaction between the $q$-mode and its thermal bath is stronger than that between the $p$-mode and its bath, resulting in the steering $S_{pq}$ dropping faster than the steering $S_{qp}$. In other words, the magnon mode with a larger damping is more difficult to steer the other one. Obviously, the one-way steering from the $p$-mode to the $q$-mode appears when the dissipation ratio satisfies $2.1<\kappa_{q}/\kappa_{p}<2.3$, where applying Gaussian measurements on the $p$-mode could unidirectionally steer the $q$-mode but not vice versa. As $\kappa_{q}/\kappa_{p}$ continues to increase, both steering $S_{qp}$ and $S_{pq}$ disappear completely, but the magnon entanglement still exists, which is attributed to the stricter quantum correlation of steering than that of entanglement.

The effective magnon number of $p$- and $q$-modes closely related to the directionality of EPR steering versus the dissipation ratio $\kappa_{q}/\kappa_{p}$ are plotted in Fig. \ref{fig5}(b). Compared with Fig. \ref{fig5}(a), it is evident that one mode with a larger population has more advantages in steering the other mode. The parameter region $N_{q}>N_{p}$ corresponds to the steering from the $q$-mode to the $p$-mode with
$S_{qp}>S_{pq}$, and the reversal of the steering direction that from the $p$-mode to the $q$-mode corresponds to $N_{q}<N_{p}$. The populations of the two modes are equal for the bidirectionally symmetric steering at $\kappa_{q}/\kappa_{p}\approx1.5$.

\begin{figure}
\includegraphics[width=9cm]{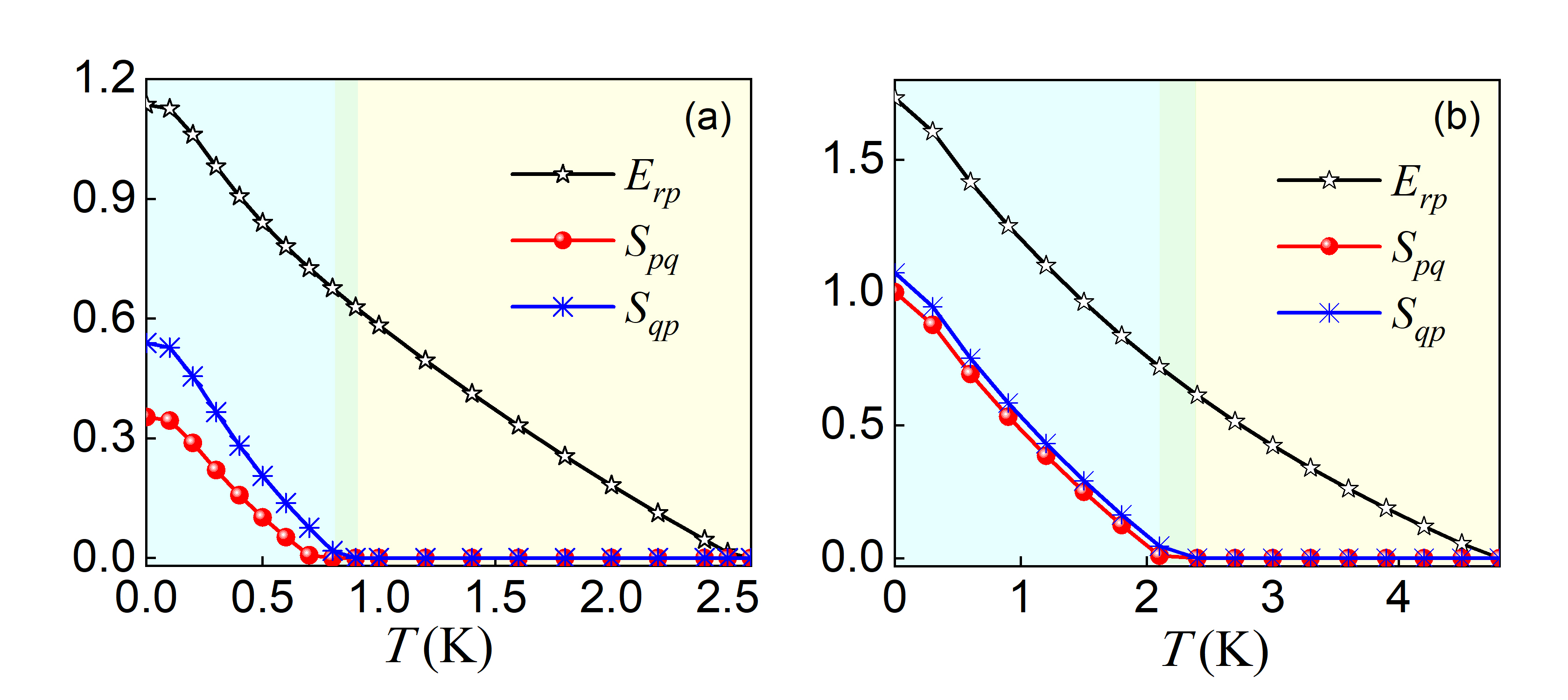}
\caption{\label{fig6} Bipartite entanglement $E_{pq}$, bipartite steering $S_{pq}$, and
$S_{qp}$ versus temperature $T$ for different effective coupling
strengths (a) $G_{p}=10G_{0}$, $G_{q}=0.85G_{p}$, and $\kappa_{r}/2\pi=40$ MHz, and (b) $G_{p}=40G_{0}$, $G_{q}=0.95G_{p}$, and $\kappa_{r}/2\pi=60$ MHz.}
\end{figure}

The entanglement and steering between $p$- and $q$-modes as functions of the environment temperature for $G_{q}=10G_{0}$ and $G_{q}=40G_{0}$ are illustrated in Fig. \ref{fig6}. It shows that the thermal noise would degrade the quantum correlation and meanwhile the steering is more sensitive to the temperature than the entanglement. From Fig. \ref{fig6}(a), we
can see that the entanglement is robust against the thermal fluctuation, which can survive up to $T\simeq2.5$ K for $G_{q}=10G_{0}$. The asymmetric steering is
reasonably robust against temperature when $T<0.9$ K, and the ability of the $q$-mode to steer the $p$-mode is always larger than the other way around (i.e., $S_{qp}>S_{pq}$). As shown in Fig. \ref{fig6}(b), the survival temperature of entanglement can be extended to $4.5$ K under a stronger effective coupling $G_{q}=40G_{0}$. Similarly, it is obvious that the degree of two-way steering and its robustness against temperature could be significantly improved, and the asymmetric steering even exists above $2$ K. Theoretically, if the resonance frequency of magnons can be increased to the THz range such as in antiferromagnets, the preparation and manipulation of entanglement can work at a temperature of several tens of Kelvin. Since the THz MFC has been theoretically predicted \cite{jin2023nonlinear,yao2023terahertz,zhang2024anisotropic}, we envision that our results can be extended to the antiferromagnetic systems  in a straightforward way.

\section{Verification and Measurement of quantum Entanglement}\label{Verification and Measurement}

\begin{figure}
\includegraphics[width=8.5cm]{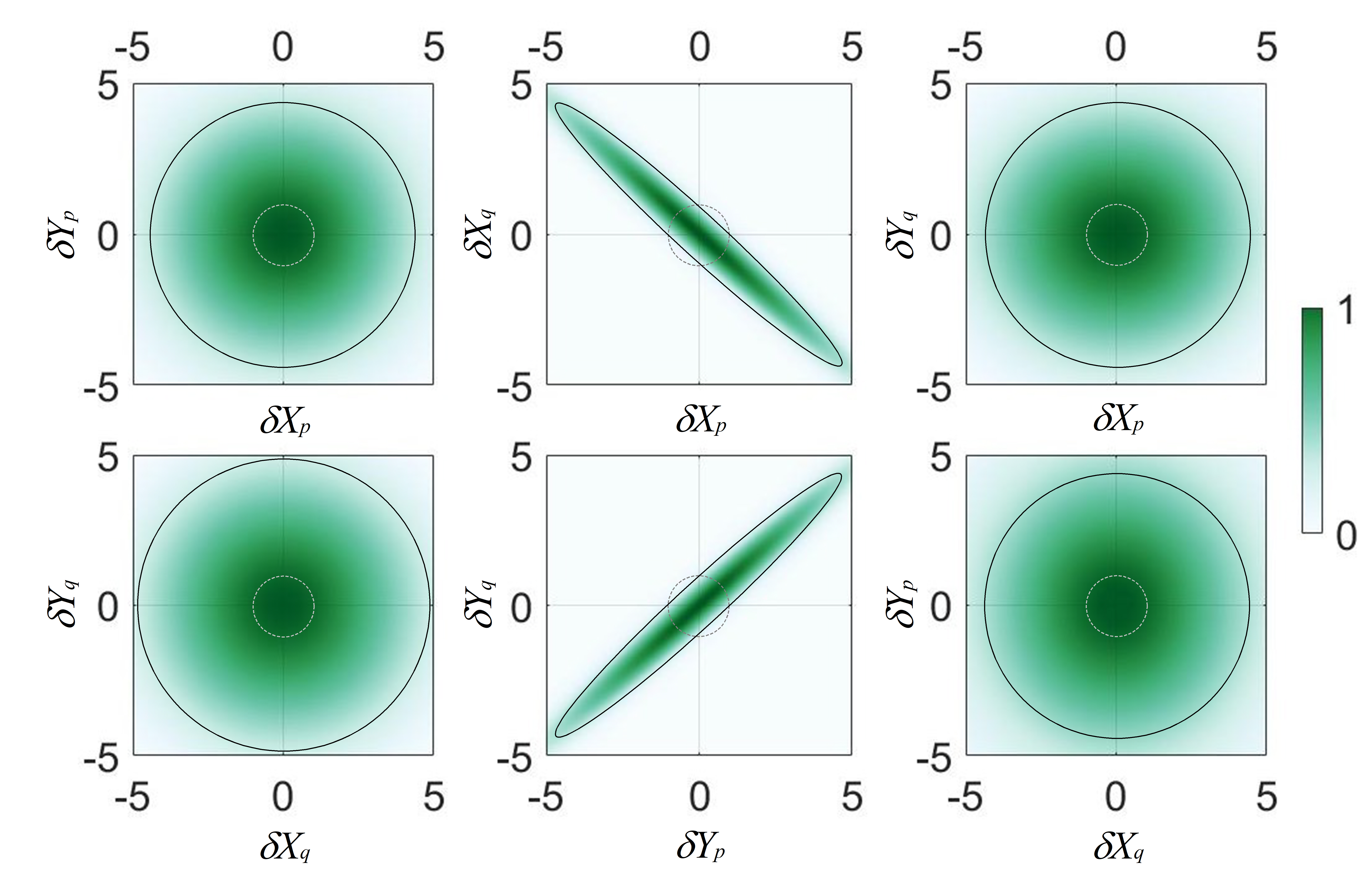}
\caption{\label{fig7} Reconstructed Wigner functions of different quadrature pairs. The solid black (dashed gray) line represents the 1/e fall-off from the maximum value of $W(\mathbf{\mu}_{pq})$  for the relevant steady state (vacuum) of the corresponding subsystem. $G_{p}=10G_{0}, G_{q}=0.85G_{p}$, $\kappa_{r}/2\pi=40$ MHz, and other parameters are in the text.}
\end{figure}

From the above analysis, it can be found that  the first-order frequency comb teeth produce two-mode squeezed entanglement under the condition of strong effective magnon-skyrmion coupling. To verify the quantum correlation, the two-mode squeezed quadratures of the sum- and difference-frequency magnon modes in the phase space are more intuitively visualized in terms of the quasiprobability Wigner functions. Because the steady state of the coupled system belongs to the Gaussian state class, its reconstructed Wigner function $W(\mathbf{\mu
}_{pq})$ can be characterized by a multivariate normal distribution, which is defined as
\begin{equation}
W(\mathbf{\mu}_{pq})=\frac{\exp(-\frac{1}{2}\mathbf{\mu}_{pq}V_{pq}%
^{-1}\mathbf{\mu}_{pq}^{T})}{\pi^{2}\sqrt{\det(V_{pq})}},\label{19}%
\end{equation}
where $\mathbf{\mu}_{pq}=(\delta X_{p},\delta Y_{p},\delta X_{q},\delta
Y_{q})$. Figure \ref{fig7} shows the reconstructed Wigner function of different quadrature pairs. Note that when the boundary of the solid black ellipse enters into the dashed gray circle, it indicates the existence of a squeezed state, where the variance of the relevant steady state is smaller than that of a vacuum state. It can be seen that the marginals from the same quadratures $(\delta X_{p},\delta Y_{p})$ and $(\delta
X_{q},\delta Y_{q})$ show unsqueezed thermal noise above the vacuum noise. The cross-quadrature marginals $(\delta X_{p},\delta X_{q})$ and $(\delta Y_{p},\delta Y_{q})$ show two-mode squeezing in the diagonal and off-diagonal directions, which indicates that two-mode squeezed correlation appear for the associated magnons.  The generated two-mode squeezing could be applied in the CV quantum processing and the quantum-enhanced measurement \cite{braunstein2005quantum,giovannetti2004quantum,cai2021quantum}.

Lastly, we discuss the experimental scheme to measure the generated quantum entanglement and steering of MFC. The quantization of bipartite quantum correlation $E_{N}$ and $S_{12(21)}$ can be obtained via measuring the elements of the CM $V$ at the steady state. The method of measuring the magnon is analogous to the mechanical oscillator. It has been experimentally realized for the case of entangled mechanical modes with the help of the microwave probe technique \cite{ockeloen2018stabilized,kotler2021direct}. As shown in Fig. \ref{fig1}(a), two weak microwave fields are sent to be resonantly coupled with magnons with different frequencies in the output MFC. The quantum state of the magnon is efficiently transferred to the microwave photon via the linear beam-splitter interaction. Therefore, the position and momentum of the magnon mode can be read by the homodyne detection of the output microwave mode. The relevant dynamics of the magnon-microwave interaction can be described by \cite{vitali2007optomechanical}
\begin{equation}
\delta\dot{c}=-\kappa_{c}\delta c-ig\delta a+\sqrt{2\kappa_{c}}\delta c^{\text{in}}, \label{20}
\end{equation}
where $\kappa_{c}$ and $\delta c^{\text{in}}$ are the dissipation and the input noise of microwave, respectively, and $g$ is the magnon-microwave coupling strength. When the microwave dissipation $\kappa_{c}\gg g$, the microwave mode adiabatically follows the magnon dynamics. From the input-output relationship \cite{gardiner1985input} $c^{\text{out}}=\sqrt{2\kappa_{c}}c-c^{\text{in}}$, we get
\begin{equation}
\delta c^{\text{out}}=\frac{-i\sqrt{2}g}{\sqrt{\kappa_{c}}}\delta a+\delta c^{\text{in}}. \label{21}
\end{equation}
It indicates that the measurement of the output microwave give the magnon dynamics, and then all elements of $V$ can be determined by measuring the correlations between the two microwave outputs, which allows one to determine both the quantum entanglement and steering. 

\section{Conclusion}\label{conclusion}
In conclusion, we have investigated the bipartite quantum correlation between different teeth of the MFC in a hybrid magnon-skyrmion system. We showed that the entanglement between the first-order comb lines can be generated, which originates from the transfer of the entanglement between the skyrmion mode and the difference-frequency magnon mode. The strong entanglement and asymmetric steering between the sum- and difference-frequency magnon modes can be obtained by controlling the magnon-skyrmion coupling strength and the skyrmion dissipation. The steering directionality could be manipulated by tuning the dissipation rates of two magnon modes, and the one-way  EPR steering can be achieved at the steady state. We found that both the bipartite entanglement and steering are robust against thermal fluctuations, and their correlation and survival temperature could be significantly improved by enhancing the driving microwave field or by extending the working frequency of MFC to the THz region. Our work reveals the fundamental quantum entanglement and EPR steering of MFC, which could have potential applications in enhancing quantum precision measurement and multiparty quantum teleportation networks. The generalization of current work to higher-order comb teeth and multipartite cases is an open issue for future study.

\begin{acknowledgments}
This work was funded by the National Key R$\&$D Program under Contract No. 2022YFA1402802 and the National Natural Science Foundation of China (NSFC) (Grants No. 12374103, No. 12434003, and No. 12074057).
\end{acknowledgments}




\end{document}